\documentstyle[11pt,epsf]{article}
\newcommand{\Par}[1]{\frac{\partial}{\partial #1}}
\newcommand{\parh}[1]{\partial^{#1}}
\newcommand{\parl}[1]{\partial_{#1}}
\newcommand{\be}{\begin{equation}}
\newcommand{\ee}{\end{equation}}
\newcommand{\bear}{\begin{eqnarray}}
\newcommand{\ear}{\end{eqnarray}}

\date{}
\begin{document}
\begin{titlepage}
\begin{flushright}
HD--THEP--94--35
\end{flushright}
\quad\\
\vspace{2cm}
\begin{center}
{\bf\LARGE Kosterlitz-Thouless phase transition}\\
\bigskip
{\bf\LARGE in the two dimensional linear $\sigma$-model}\\
\vspace{1cm}
M. Gr\"ater and C. Wetterich\\
\bigskip
{\em Institut f\"ur Theoretische Physik, Universit\"at Heidelberg,}\\
{\em Philosophenweg 16, 69120 Heidelberg}\\
\vspace{3cm}
{\bf Abstract:} \\
\parbox[t]{\textwidth}{
We investigate the $O(N)$ symmetric linear $\sigma$-model in two
dimensions by
means of an exact nonperturbative evolution equation. The perturbative
infrared divergences are absent in this formulation. We use a simple
approximative solution of the flow equation which corresponds to a
derivative expansion for the effective action. For $N=2$ this gives a good
picture of the Kosterlitz-Thouless phase transition.
}
\end{center}
\end{titlepage}
\newpage

In dimensions smaller than four the $N$-component linear $\sigma$-model
is not directly accessible to perturbation theory. The obstacle arises
from infrared divergences due to massless modes - either the Goldstone
modes in the regime with spontaneous symmetry breaking for $N\geq 2$ or
the massless mode at the phase transition. One usually
ressorts to an extrapolation of the $4-\epsilon$ expansion around the
four dimensional model, or to the $2+\epsilon$ expansion around the
corresponding two dimensional nonlinear $\sigma$-model. We present
here an approach working in arbitrary dimension $d$ which can cope
with the infrared problems. It is based on the concept of the average
action $\Gamma_k$ \cite{A} which is the continuum analogue of the
block spin action on the lattice \cite{B}. The computation of
$\Gamma_k$ includes all fluctuations with momenta $q^2$ larger than
the infrared cutoff $k^2$. For $k\to 0$ one recovers the effective
action, i.e. the generating functional for the one particle
irreducible Green functions or the free energy. The dependence of
$\Gamma_k$ on the scale $k$ is determined by an exact evolution
equation \cite{cw1}:
\be
\label{master}
\Par{t}\Gamma_{k}=\frac{1}{2}Tr\left\{(\Gamma^{(2)}_{k}+R_{k})^{-1}
\Par{t}R_{k}
\right\}.
\ee
(In a Fourier basis the trace reads
$Tr=\sum_{a=1}^{N}\int \frac{d^dq}{(2\pi)^d}$ and $t=\ln (k/\Lambda)$.)
In distinction to earlier versions of exact renormalization group
equations \cite{C} the flow equation (\ref{master}) is
characterized by its close resemblance to perturbation theory:
The main differences as compared to a one loop expression are the
appearance of the exact inverse propagator $\Gamma_k^{(2)}$
(the second functional variation of $\Gamma_k$ with respect to
the $N$-component scalar fields $\phi_a$) and the infrared
cutoff $R_k$.
With
\be
\label{Rk}
R_{k}(q^2)= \frac{Z_{k}q^{2}\exp
(-q^{2}/k^{2})}{1-\exp (-q^{2}/k^{2})}
\ee
(where $Z_k$ is an appropriate wave function renormalization
constant specified below) we observe that the momentum integral in
(\ref{master}) is both infrared and ultraviolet finite. The short distance
physics has to be specified by the ``initial value'' $\Gamma_\Lambda$ (the
``classical action'') at some high momentum scale $\Lambda$. A solution of
the flow equation (\ref{master}) allows to extrapolate from the classical
action $\Gamma_\Lambda$ to the effective action $\Gamma_0$ and therefore
constitutes a solution of the model.

Our nonperturbative evolution equation remains a complicated nonlinear
functional differential equation which cannot be solved exactly.
One needs approximate nonperturbative solutions which correspond to a
truncation of the general form of $\Gamma_k$ and therefore reduce
the problem to a manageable number of degrees of freedom. We consider
here the $O(N)$ symmetric linear $\sigma$-model with a truncation
corresponding to the lowest terms in a derivative expansion of $\Gamma_k$,
i.e.
\be
\label{derexp}
\Gamma_{k}=\int d^{d}x \left\{ U_{k}(\rho )+\frac{1}{2}Z_{k}
\parh{\mu} \phi_{a} \parl{\mu}\phi^{a} \right\},
\ee
where $\rho=\frac{1}{2} \phi_a\phi^a$.
The truncation (\ref{derexp}) neglects all terms with
more than two derivatives.
The most general two derivative expression would contain for
$N>1$ a second term
$Y_k\parl{\mu}\rho\parh{\mu}\rho $ and $Z_k$ and $Y_k$ could depend
on $\rho$. For simplicity we consider first only a constant $Z_k$ and
$Y_k=0$.

With the Ansatz (\ref{derexp}) we obtain from (\ref{master}) an evolution

equation for the effective average potential:

\be
\label{dimpotevo}
\Par{t}U_k(\rho)=v_d\int_0^\infty dx x^{d/2} s_k(x)
\left( \frac{N-1}{M_0}+\frac{1}{M_1}\right)
\ee
where $x=q^2$ with
\bear
M_0&=&x(1+r_k(x))+Z_k^{-1}U'_k(\rho)\\
M_1&=&x(1+r_k(x))+Z_k^{-1}U'_k(\rho)+2 Z_k^{-1}\rho U''_k(\rho)
\nonumber\\
v_d^{-1}&=&2^{d+1}\pi^{d/2} \Gamma (d/2)\nonumber.
\ear
Here primes denote derivatives with respect to $\rho$ and we observe
the appearance of $\rho$ dependent  mass terms $\sim U'_k$ in the
``Goldstone'' and $\sim U'_k+2\rho U''_k$ in the ``radial'' directions.
The dimensionless functions $s_k$ and $r_k$ depend on the ratio $x/k^2$
and are given by
\bear
\label{runds}
r_{k}(x)&=&\frac{R_{k}(x)}{Z_{k}x}\nonumber,
\\
s_{k}(x)&=&Z_{k}^{-1}\Par{t}\left(\frac{R_{k}(x)}{x}\right)
=-2x\Par{x}r_{k}(x)-\eta r_k(x).
\ear
The second term in $s_k(x)$ is proportional to the anomalous
dimension
\be
\label{etadef}
\eta =-\Par{t}\ln Z_k
\ee
and arises from the factor $Z_k$ in (\ref{Rk}).
We will neglect it in the following.

\noindent
It is convenient to work with dimensionless and renormalized
quantities:
\be
\tilde\rho =k^{2-d}Z_k\rho  \quad\mbox{;}\quad
u_k(\tilde\rho)=k^{-d}U_k\left(\rho(\tilde\rho)\right) \quad\mbox{;}
\quad y=\frac{x}{k^2}.
\ee
To evaluate the equation for the potential we make a further approximation
and expand around the minimum of $u_k$ for non zero $\tilde\rho=\kappa$ up
to the
quadratic order in $\tilde\rho$:
\be
\label{potentw}
u_k(\tilde\rho)= u_k(\kappa)+\frac{1}{2}\lambda (\tilde\rho-\kappa)^2.
\ee
The condition $\partial u/\partial\tilde\rho\vert_{\tilde\rho=\kappa}=0$
holds independent of $t$ and gives us the evolution equation
for the location of the minimum of the potential parametrized by $\kappa$.
A similar evolution equation can be derived \cite{tw1} for the
symmetric regime where $\kappa=0$ and an appropriate variable is
$\partial u /\partial \tilde\rho\vert_{\tilde\rho=0}$. The flow equations
for $\kappa$ and $\lambda$ follow from (\ref{dimpotevo}) by variable
substitution
and appropriate differentiations with respect to $\tilde\rho$:
\bear
\label{betakappa}
\beta_\kappa &\equiv& \frac{d\kappa}{dt}  = -(d-2+\eta)\kappa +
2 v_d(N-1) l^d_1(0)
+6 v_d  l^d_1 (2\lambda\kappa)
\nonumber\\
\label{betalambda}
\beta_\lambda &\equiv& \frac{d\lambda}{dt} = (d-4+2\eta)\lambda
+2v_d(N-1)\lambda^{2} l^d_2(0)
+18 v_{d}\lambda^{2} l^d_2(2\lambda\kappa).
\ear
We notice that the ``threshold functions''
\bear
\label{lintegrals}
l^d_n(\omega)&=& \frac{n}{2} \int^{\infty}_{0} dy y^{\frac{d}{2}}
s(y) [y(1 + r(y)) +\omega]^{-(n+1)}
\ear
vanish for $\omega\to \infty$ and describe appropriately the decoupling
of a heavy radial mode for $2\lambda\kappa \gg 1$.
The integrals are normalized such that $l^2_1(0)=1,\: l^4_2(0)=1$.
Eq. (\ref{betalambda}) describes how the average potential
$U_k$ changes its shape, with ``initial condition'' given at some
short distance scale $k=\Lambda$ by $\kappa(\Lambda)$ and
$\lambda(\Lambda)$.
The solution for $k=0$ determines the 1PI two and four point
functions at zero momentum. In particular, the theory is in the
symmetric phase if $\kappa(0)=0$ - this happens if $\kappa$ reaches zero
for some nonvanishing $k_s>0$. On the other hand, the phase with
spontaneous symmetry breaking corresponds to $\rho_0(0)>0$ where
$\rho_0(k)=k^{d-2}Z_k^{-1}\kappa(k)$. A second order phase transition is
characterized by a scaling solution corresponding to fixpoints
for $\kappa$ and $\lambda$. For small deviations from the fixpoint there
is
typically one
infrared unstable direction which is related to the relevant mass
parameter.
The phase transition can be studied as a function of $\kappa(\Lambda)$
with a critical value $\kappa(\Lambda)=\kappa_c$. The difference
$\kappa(\Lambda)-\kappa_c$ can be assumed to be proportional to $T_c-T$,
with $T_c$ the critical temperature. This allows to define and compute
critical exponents in a standard way. We should mention a particular
possibility for $d=2$, namely that $\kappa(0)$ remains strictly
positive whereas $\rho_0(0)$ vanishes due to $\lim_{k\to 0}Z_k\to\infty$.
This is a somewhat special form of spontaneous symmetry breaking, where
the renormalized expectation value, which determines the renormalized
mass, is different from zero whereas the expectation value of the
unrenormalized field vanishes. We will see that this scenario is indeed
realized for $d=2, N=2$. The phase with this special form of spontaneous
symmetry breaking exhibits a massive radial and a massless Goldstone
boson - and remains nevertheless consistent with the Mermin-Wagner theorem
\cite{merwag} that the expectation value of the (unrenormalized) field
$\phi_a$ must vanish for $N\ge2$. The Kosterlitz-Thouless phase transition
\cite{kostthou} describes the transition from this phase to the standard
symmetric phase of the linear $\sigma$-model, i.e. the phase where
$\kappa(0)=0$ with a spectrum of two degenerate massive modes.

In order to solve the flow equation (\ref{betalambda}) we further need
the anomalous dimension $\eta$ (\ref{etadef}) which is related to the
scale dependence of the wave function re\-nor\-ma\-li\-zation.
We identify the wave function renormalization in the infrared cutoff
(\ref{Rk}) with the coefficient multiplying the kinetic term in the
average action (\ref{derexp}), evaluated for zero momentum and
$\tilde\rho=\kappa$. The flow of $Z_k$ can then be computed from
the momentum dependence of $\Par{t}\Gamma^{(2)}_k$
\cite{A} and one obtains with the truncation (\ref{derexp})
\bear
\label{evowfr}
\eta=\frac{16 v_d}{d}\lambda^2\kappa\: m^d_{2,2}\left(0,2
\lambda\kappa\right).
\ear
The integral $m^d_{2,2}$ defines again an appropriate threshold function
\bear
\label{mintegrals}
m^{d}_{2,2 }(0,\omega)=-\frac{1}{2}\int_0^{\infty} dy
y^{d/2}\widehat{\Par{t}} \left\{ \left( 1+r(y)+y\Par{y}r(y)\right)^2
[y(1+r(y))]^{-2}[y(1+r(y))+\omega]^{-2}\right\}.
\ear
Here $\widehat{\Par{t}}$ stands symbolically for a derivative which
acts only on the infrared cutoff $R_k$ in $r$, with
$\widehat{\Par{t}}r(y)\equiv s(y)$, and we omit again the second term
$\sim\eta$
in $s$ (\ref{runds}). We have now a system of three coupled
non linear differential equations for $\kappa,\lambda$ and
$\eta$ which can be solved numerically for arbitrary $N$ and $d$.

We next specialize to the two dimensional linear $\sigma$-model ($d=2$).
We observe that the flow equations can be solved analytically in the
limiting case
of a large mass $\omega=2\lambda\kappa$ of the radial mode.
The threshold functions vanish with powers of $\omega^{-1}$ and for
$N>1$ the leading contributions to the $\beta$-functions are those
from the Goldstone modes. Therefore this limit is called the Goldstone
regime. In this approximation the $\beta$-functions can be expanded
in powers of $\omega^{-1}$. In particular, the leading order of the
anomalous dimension can be extracted immediately from (\ref{evowfr}):
\be
\label{etaentw}
\eta = \frac{1}{4\pi\kappa}+{\em O} (\kappa^{-2}).
\ee
Inserting this result in (\ref{betakappa}) we have
\be
\label{kappaent}
\beta_\kappa= \frac{(N-2)}{4\pi} + {\em O}(\kappa^{-1})
\ee
and the leading order of $\beta_\lambda$ is
\be
\label{entbetala}
\beta_\lambda = -2\lambda+\frac{(N-1)\ln 2}{2\pi}\lambda^2  + {\em
O}(\kappa^{-1}).
\ee
Eq. (\ref{entbetala}) has a fixpoint solution
$\lambda_*=\frac{4\pi}{(N-1)\ln 2}
\approx 18.13/(N-1)$.

For $N>2$ there exists a simple relation between the linear and
the nonlinear $\sigma$-model:
The effective coupling between the Goldstone bosons of the nonabelian
nonlinear $\sigma$-model can be extracted directly from (\ref{derexp})
and reads in an appropriate normalization \cite{LWW}
\be
g^2=\frac{1}{2\kappa}.
\ee
The lowest order contribution to $\beta_\kappa$ (\ref{kappaent})
coincides with the one loop expression for the running of $g^2$ as
computed in the nonlinear $\sigma$-model. We emphasize in this
context the importance of the anomalous dimension $\eta$ which changes
the factor $(N-1)$ appearing in (\ref{betalambda}) into the appropriate
factor $(N-2)$ in (\ref{kappaent}). In correspondence with the
universality of the two loop $\beta$-function for $g^2$ in the nonlinar
$\sigma$-model we expect the next to leading term $\sim\kappa^{-1}$
in $\beta_\kappa$ (\ref{kappaent}) to be also proportional to
$(N-2)$. In order to verify this one has to go beyond the truncation
(\ref{derexp}) and systematically keep all terms contributing in the
appropriate order of $\kappa^{-1}$. (This calculation is similar
to the extraction of the two loop $\beta$-function of the linear
$\sigma$-model in four dimensions by means of an ``improved one
loop calculation'' using the flow equation (\ref{master}) \cite{pw}.)
We have calculated the expansion of $\beta_\kappa$ up to
the order ${\em O}(\kappa^{-1})$ for the most general two derivative
action, i.e. neglecting only the momentum
dependence of $Z_k$ and $Y_k$. The result agrees with the
two loop term of the nonlinear $\sigma$-model within a few
per cent, and the discrepancy should be attributed to the neglected
momentum dependence of the wave function renormalization. The issue
of the contribution to $\beta_\kappa$ in order $\kappa^{-2}$ is less
clear:
Of course, the direct contribution of the Goldstone bosons
(combined with their contribution to $\eta$) should always vanish
for $N=2$ since no nonabelian coupling exists in this case. The
radial mode however, could generate a contribution which is not
proportional to $(N-2)$. This contribution is possibly nonanalytic
in $\kappa^{-1}$ and would correspond to a nonperturbative
contribution in the language of the nonlinear $\sigma$-model.

Let us now turn to the two dimensional abelian model ($d=2, N=2$)
for which we want to describe the Kosterlitz-Thouless phase transition.
In the limit of vanishing $\beta_\kappa$ for large enough $\kappa$ the
location of the minimum of $u_k(\tilde\rho)$ (\ref{potentw}) is
independent of the scale $k$. Therefore the
parameter $\kappa$, or, alternatively, the temperature difference
$T_c-T$, can be viewed as a free parameter.
If we go beyond the lowest order estimate (\ref{entbetala})
the fixpoint for $\lambda$ remains, but $\lambda_*$ becomes
dependent on $\kappa$. This implies that the system has
a line of fixpoints which is parametrized by $\kappa$ as
suggested by results obtained from calculations with the
nonlinear $\sigma$-model \cite{webe}.
In particular, the anomalous dimension $\eta$ depends on the
temperature $T_c-T$ (\ref{etaentw}). Even if this picture is not
fully accurate for nonvanishing $\beta_\kappa$, it is a very
good approximation for large $\kappa$: The possible running
of $\kappa$ is extremely slow, especially if $\beta_\kappa$
vanishes in order $\kappa^{-1}$.
We associate the low temperature or large $\kappa$ phase with
the phase of vortex condensation in the nonlinear $\sigma$-model.
The correlation lenght is always infinite due to the Goldstone boson.
Since $\eta>0$ we expect the inverse propagator of this Goldstone
degree of freedom $\sim (q^2)^{1-\eta/2}$, thus avoiding Coleman's
no go theorem \cite{cole} for free massless particles in two dimensions.
On the other hand, for small values of $\lambda\kappa$ the threshold
functions (\ref{lintegrals})(\ref{mintegrals}) can be expanded in
powers of $\lambda\kappa$. The anomalous dimension is small and
$\kappa$ is driven to zero for $k_s>0$. This corresponds to the symmetric
phase of the linear $\sigma$-model with a massive complex scalar field.
We associate this high temperature phase with the phase of vortex disorder
in the picture of the nonlinear $\sigma$-model.
The transition between the behaviour for large and small $\kappa$ is
described by the Kosterlitz-Thouless transition. In the language of the
linear $\sigma$-model it is the transition from a special type of
spontaneous symmetry breaking to symmetry restoration.

Finally we give a summary of the results obtained from the
numerical integration of the evolution equations (\ref{betalambda})
and (\ref{evowfr}) for the special
case $d=2, N=2$ \footnote{This work has also been done for $d=2, N=1$.
There we find a fixpoint which corresponds to the second
order phase transition in the Ising model.}.
We use a Runge-Kutta method starting at $t=0$ with arbitary initial
values for $\kappa$ and $\lambda$ and solve the flow equations for
large negative values of $t$.
For the integrals defined by (\ref{lintegrals}) and
(\ref{mintegrals}) we use numerical fits.
Results are shown in figs.1-4 where we plot typical trajectories.
The distance between points corresponds to equal steps in $t$ such
that very dense points or lines indicate the very slow running in the
vicinity
of fixpoints.

The understanding of the trajectories needs a
few comments:
The work of Kosterlitz and Thouless \cite{kostthou} suggests that the
correlation lenght is divergent for all temperatures below
a critical temperature $T_c$ and that the critical exponent $\eta$ depends
on temperature. The consequence for our model
is that above a critical value for $\kappa$ all $\beta$-
functions should vanish for a line of fixpoints parametrized by $\kappa$.
{}From the results in the Goldstone
regime and from earlier calculations \cite{webe} we conclude that
$\beta_\kappa$ should vanish faster than $\kappa^{-1}$ for large $\kappa$.
Our truncation (\ref{derexp}), however,
yields a function $\beta_\kappa$ which vanishes only like $\kappa^{-1}$.
The consequence is that even if the system reaches the supposed line of
fixpoints the parameter $\kappa$ decreases very slowly until the
transition to the symmetric regime is reached.
The anomalous dimension first grows with decreasing $\kappa$
(\ref{etaentw})
until the critical value is reached. Then the
system runs into the symmetric regime and $\eta$ vanishes. So
we expect that $\eta$ reaches a maximum  near the phase transition.
We use this as a criterion for the critical value $\kappa_c$.
In summary, the truncation (\ref{derexp}) smoothens the phase
transition and this prevents a very accurate determination of the critical
value $\kappa_c$ and the corresponding anomalous dimension $\eta_c$.
{}From the numerical point of view the absence of a true phase transition
in the
truncation (\ref{derexp}) makes life easier: One particular trajectory can
show both the features of the low and the high temperature phase since
it crosses from one to the other.

\begin{figure}[b]
\leavevmode
\centering
\epsfxsize=11cm
\epsffile{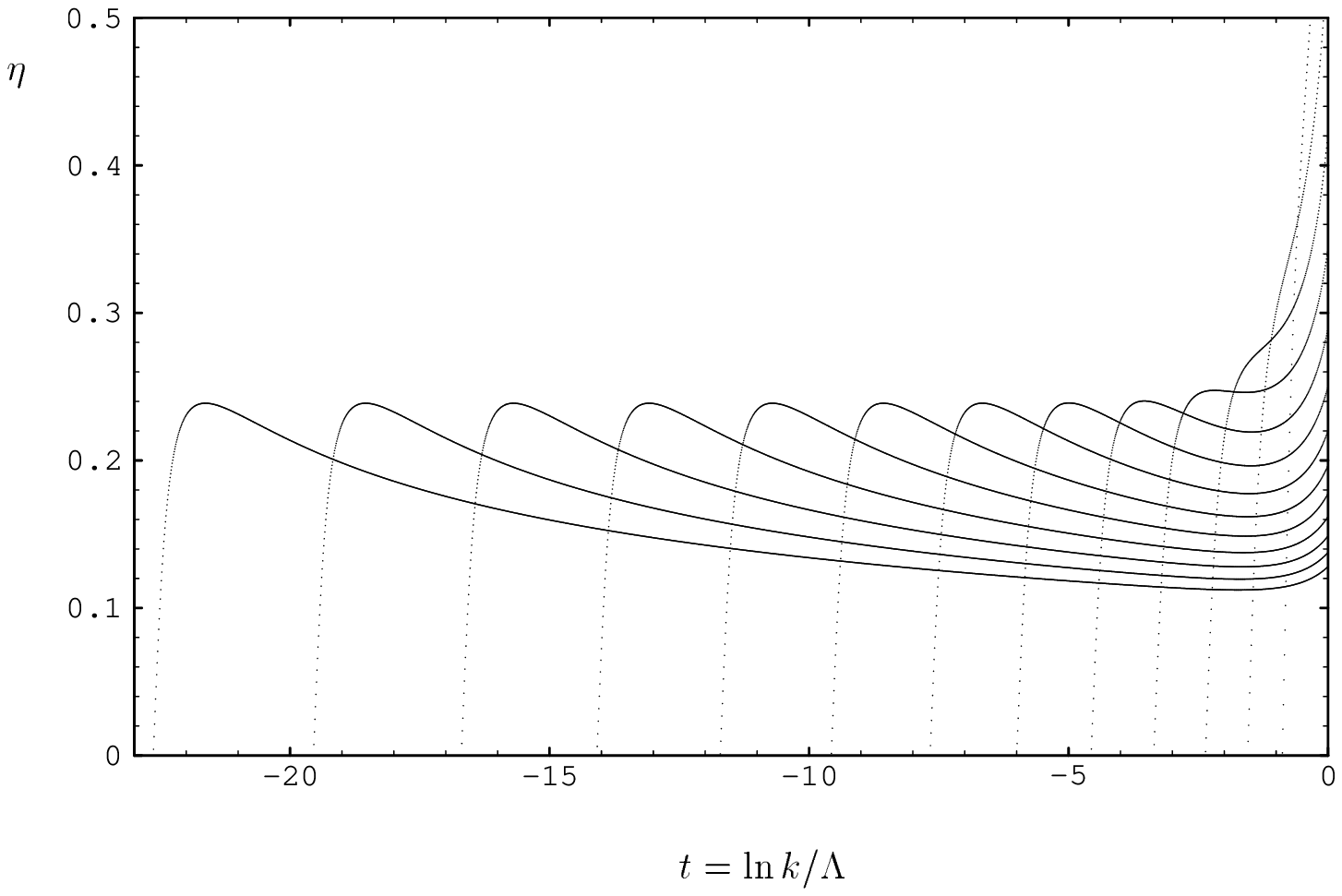}
\caption[The scale dependent anomalous dimension
$\eta$]{\label{n2etagraf} $\eta(t)$}
\end{figure}

\begin{figure}[t]
\leavevmode
\centering
\epsfxsize=11cm
\epsffile{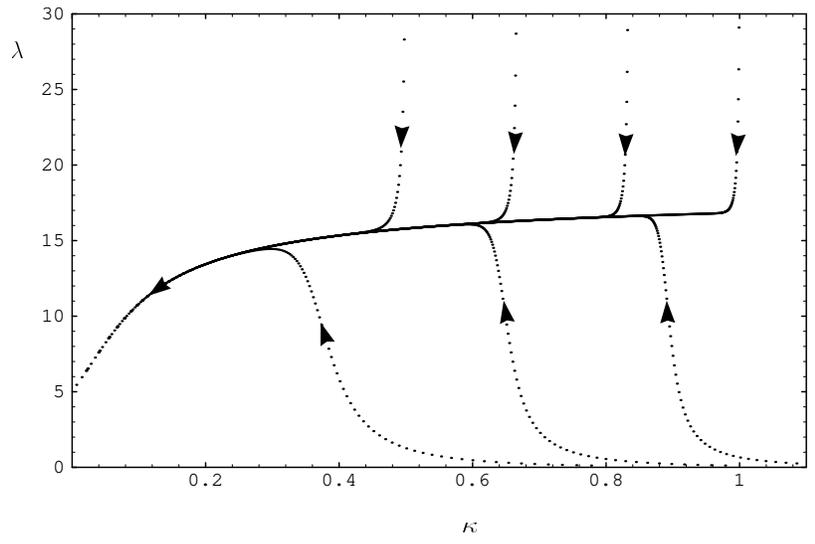}
\caption[Two dimensional parameter space]{\label{n2kalagraf}
$\lambda(\kappa)$}
\end{figure}

\begin{figure}[t]
\leavevmode
\centering
\epsfxsize=11cm
\epsffile{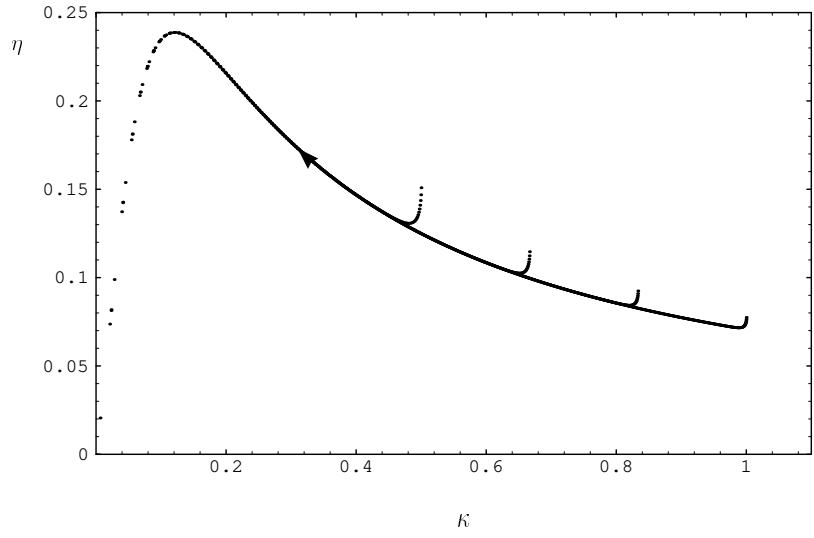}
\caption[The $\kappa$ dependence of the critical exponent
$\eta$]{\label{n2etakagraf} $\eta(\kappa)$}
\end{figure}

\begin{figure}[b]
\leavevmode
\centering
\epsfxsize=11cm
\epsffile{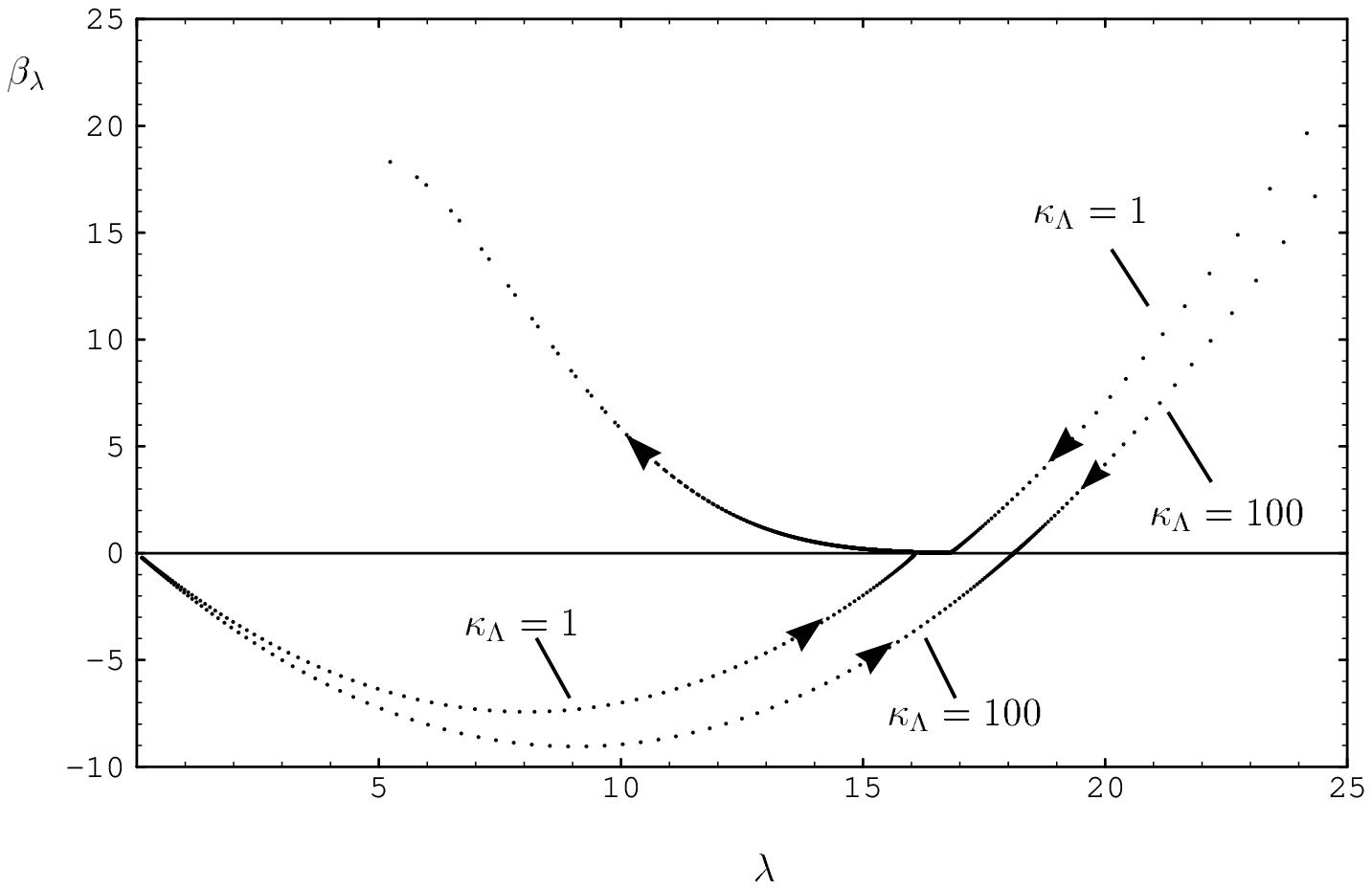}
\caption[The $\beta$-function for $\lambda$]
{\label{n2belagraf} $\beta_\lambda $}
\end{figure}

The numerical results fullfil our expectations. Fig.\ref{n2etagraf}
shows the evolution of the anomalous dimension with decreasing $t$ for
several different initial values $\kappa(\Lambda), \lambda(\Lambda)$.
The maximum is reached with $\eta_c=0.24$ which has to be compared
with the result of Kosterlitz and Thouless $\eta_c=0.25$ \cite{kostthou}.
The approximate ``line of fixpoints'' for $\kappa>\kappa_c$
$(\eta<\eta_c)$
is demonstrated by the selfsimilarity of the curves for large $-t$.
Trajectories with different initial conditions hit the line of fixpoints
at
different $\kappa$. Subsequently they follow the line of fixpoints. Except
for the value of $\kappa(k)$ all ``memory'' of the initial conditions is
lost
for $t\le-3$. Another manifestation of the line of fixpoints in the
$(\kappa,\lambda)$ plane is demonstrated in fig.\ref{n2kalagraf}. After
some
fast ``initial running'' (dotted parts of the trajectories) all
trajectories
with large enough $\kappa(\Lambda)$ follow this line independent of the
initial
$\lambda(\Lambda)$. We emphasize that the nonlinear $\sigma$-model
corresponds to $\lambda(\Lambda)\to\infty$. Our investigation shows that
the
linear $\sigma$-model is in the same universality class, even for very
small
$\lambda(\Lambda)$. In fig.\ref{n2etakagraf} we plot $\eta(\kappa)$. Along
the line of fixpoints we find perfect agreement with the analytical
estimate
(\ref{etaentw}) for large $\kappa$. For $\kappa=0.3$ the deviation from
the
lowest order result is 34\% in the present truncation. Finally we show in
fig.\ref{n2belagraf} the value of $\beta_\lambda$ for different
trajectories.
The dense parts of the ``ingoing curves'' show the fixpoint behaviour at
$\lambda_*(\kappa_1)$ where $\kappa_1$ denotes the value of $\kappa$ where
the line of fixpoints is hit. (For small $\kappa(\Lambda)$ there is a
substantial difference between $\kappa_1$and $\kappa(\Lambda)$ which
depends
also on $\lambda(\Lambda)$. This can be seen from the curves with
$\kappa(\Lambda)=1$.) After hitting the line of fixpoints the trajectories
stay for a large $t$-interval at $\beta_\lambda$ very close to zero.
Subsequently, the ``outgoing curve'' indicates the transition to the
symmetric phase.

In conclusion, both the analytical and the numerical investigations
demonstrate
all important characteristics of the Kosterlitz-Thouless phase transition
for the linear $\sigma$-model. This belongs to the same universality class
as the nonlinear $\sigma$-model and we have demonstrated a close
correspondence
between the linear and the nonlinear $\sigma$-model with abelian symmetry.
In
particular, the phase with vortex disorder in the nonlinear $\sigma$-model
corresponds simply to the symmetric phase of the linear $\sigma$-model. We
emphasize that we have never needed the explicit investigation of vortex
configurations. The exact nonperturbative flow equation includes
automatically
all configurations. Its ability to cope with the infrared problems of
perturbation theory is confirmed by the present work.

Despite the simple and clear qualitative picture arising from the
truncation
(\ref{derexp}) this letter only constitutes a first step for a
quantitative
investigation. It is not excluded that the coincidence of our critical
$\eta_c\approx 0.24$ with $1/4$ is somewhat accidental. In order to answer
this question one needs to go beyound the truncation (\ref{derexp}). In
view
of the relatively large value of $\eta_c$ we expect in particular that the
momentum dependence of the wave function renormalization $Z_k$ (or the
deviation of the inverse propagator from $q^2$) could play an important
role
at the phase transition. This effect should be included in a more detailed
quantitative investigation.

\newpage
\section*{Figure Captions}
\begin{itemize}
\item fig 1: The scale dependent anomalous dimension $\eta$
\item fig 2: Critical line in the $\kappa-\lambda$ plane
\item fig 3: The $\kappa$ dependence of the critical exponent $\eta$
\item fig 4: The $\beta$-function for $\lambda$
\end{itemize}
\end{document}